\documentclass[preprint,aps]{revtex4}
\usepackage{graphicx}
\usepackage{booktabs}
\usepackage[usenames, dvipsnames]{color}
\usepackage{multirow}
\usepackage{setspace}
\usepackage[ pdftex, plainpages = false, pdfpagelabels, 
                 pdfpagelayout = useoutlines,
                 bookmarks,
                 bookmarksopen = true,
                 bookmarksnumbered = true,
                 breaklinks = true,
                 linktocpage=all,
                 pagebackref=false,
                 colorlinks = true,
                 linkcolor = BrickRed,
                 urlcolor  = blue,
                 citecolor = BrickRed,
                 anchorcolor = green,
                 hyperindex = true,
                 hyperfigures
                 ]{hyperref} 
                 
\newcommand{\comment}[1]{}

\begin{document}

\title{The Food Crises and Political Instability \\
in North Africa and the Middle East} 
\date{July 19, 2011; revised August 10, 2011}
\author{Marco Lagi, Karla Z. Bertrand and Yaneer Bar-Yam}
\affiliation{New England Complex Systems Institute \\ 
238 Main St., Suite 319, Cambridge, MA 02142, USA}

\begin{abstract}
Social unrest may reflect a variety of factors such as poverty, unemployment, and social injustice. Despite the many possible contributing factors, the timing of violent protests in North Africa and the Middle East in 2011 as well as earlier riots in 2008 coincides with large peaks in global food prices. 
We identify a specific food price threshold above which protests become likely. 
These observations suggest that protests may reflect not only long-standing political failings of governments, but also the sudden desperate straits of vulnerable populations. If food prices remain high, there is likely to be persistent and increasing global social disruption. 
Underlying the food price peaks we also find an ongoing trend of increasing prices. We extrapolate these trends and identify a crossing point to the domain of high impacts, even without price peaks, in 2012-2013. This implies that avoiding global food crises and associated social unrest requires rapid and concerted action. 
\end{abstract}

\maketitle

In 2011 protest movements have become pervasive in countries of North Africa and the Middle East. These protests are associated with dictatorial regimes and are often considered to be motivated by the failings of the political systems in the human rights arena \cite{Zakaria2011,Mason2011,Shah2011,Beheshtipour2011}.  
Here we show that food prices are the precipitating condition for social unrest \cite{Arezki2011,MacKenzie2011Feb,Egypt_control3,Fraser2011,McCann2011,Kudlow2011,Bellemare2011, Brinkman2010} and identify a specific global food price threshold for unrest.
Even without sharp peaks in food prices we project that, within just a few years, the trend of prices will reach the threshold.
This points to a  danger of spreading global social disruption.  

Historically, there are ample examples of ``food riots," with consequent challenges to authority and political change, notably in the food riots and social instability across Europe in 1848, which followed widespread droughts \cite{Dowe2001}. While many other causes of social unrest have been identified, food scarcity or high prices often underlie riots, unrest and revolutions \cite{Rude1964, Quinault1974, Walton1994, Gurr1970, Tilly1978, Humphreys2007, Blattman2010}. Today, many poor countries rely on the global food supply system and are thus sensitive to global food prices \cite{Senauer2005}. This condition is quite different from the historical prevalence of subsistence farming in undeveloped countries, or even a reliance on local food supplies that could provide a buffer against global food supply conditions. It is an example of the increasingly central role that global interdependence is playing  in human survival and well-being \cite{Bar-Yam2002,MacKenzie2008Apr,MacKenzie2008Oct}. We can understand the appearance of social unrest in 2011 based upon a hypothesis that widespread unrest does not arise from long-standing political failings of the system, but rather from its sudden perceived failure to provide essential security to the population. In food importing countries with widespread poverty, political organizations may be perceived to have a critical role in food security. Failure to provide security undermines the very reason for existence of the political system. Once this occurs, the resulting protests can reflect the wide range of reasons for dissatisfaction, broadening the scope of the protest, and masking the immediate trigger of the unrest.

Human beings depend on political systems for collective decision making and action and their acquiescence to those systems, if not enthusiasm for them, is necessary for the existence of those political systems. The complexity of addressing security in all its components, from protection against external threats to the supply of food and water, is too high for individuals and families to address themselves in modern societies \cite{DCS}. 
\begin{figure}[tb]
\centerline{
	\includegraphics[width=150mm]{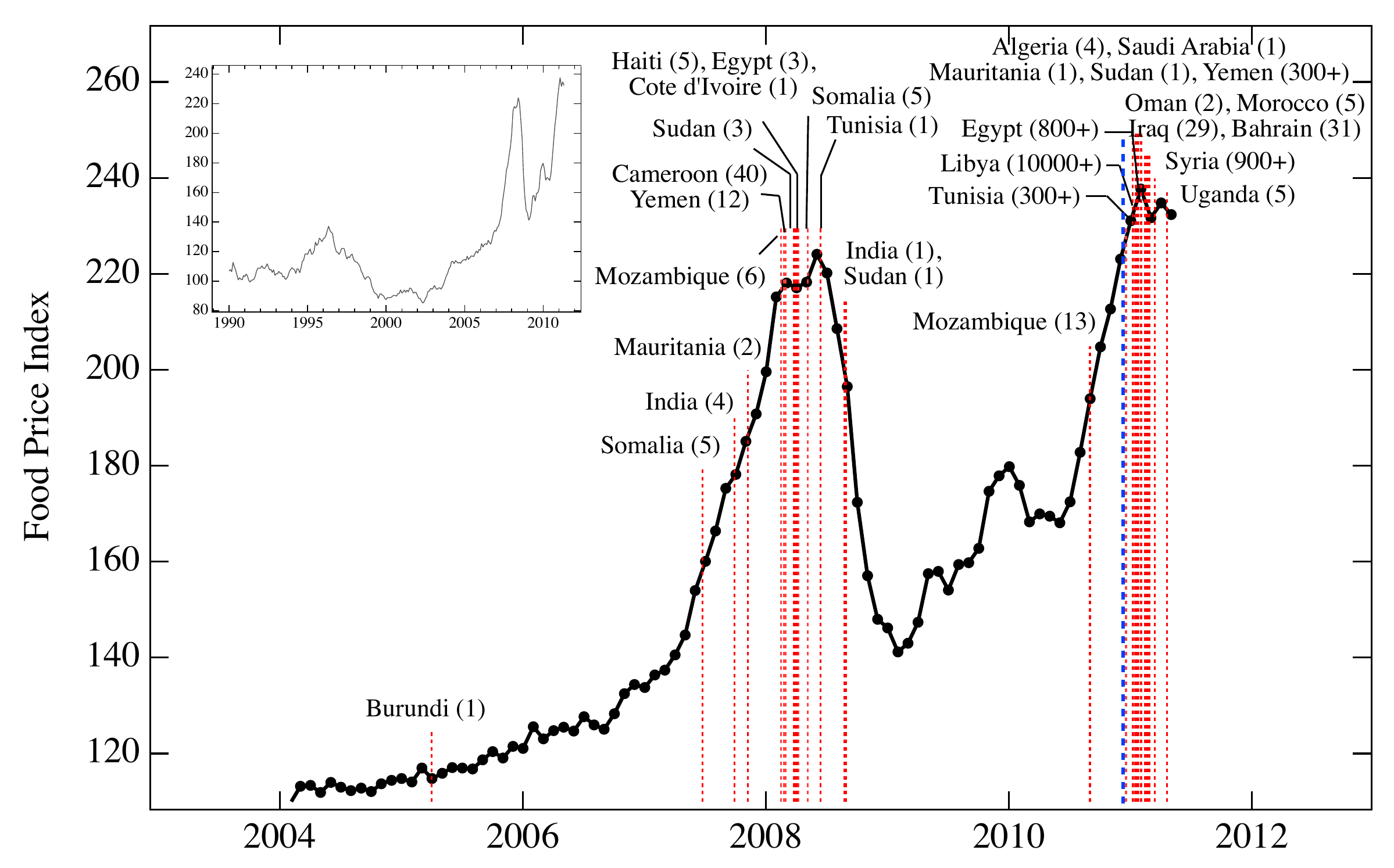}
}
\caption{Time dependence of FAO Food Price Index from January 2004 to May 2011. Red dashed vertical lines correspond to beginning dates of ``food riots" and protests associated with the major recent unrest in North Africa and the Middle East. The overall death toll is reported in parentheses \cite{Burundi1, Somalia1,India1, Mauritania1, Cameroon1, Mozambique1, Yemen1, Egypt1, Sudan1, Haiti1, CotedIvoire1, Somalia2, Tunisia1, India2, Sudan2, Mozambique2, Tunisia2, Algeria1, Egypt2, Mauritania2, Sudan3, SaudiArabia1, Morocco1, Oman1, Libya1, Yemen2, Iraq1, Bahrain1, Syria1, Uganda1}. Blue vertical line indicates the date, December 13, 2010, on which we submitted a report to the U.S. government, warning of the link between food prices, social unrest and political instability \cite{NECSIreport}. Inset shows FAO Food Price Index from 1990 to 2011.\label{fig:foodprotestsdate}}
\end{figure}
Thus, individuals depend on a political system for adequate decision making to guarantee expected standards of survival. This is particularly true for marginal populations, i.e. the poor, whose alternatives are limited and who live near the boundaries of survival even in good times. The dependence of the population on political systems engenders its support of those systems, even when they are authoritarian or cruel, compromising the security of individuals while maintaining the security of the population. Indeed, a certain amount of authority is necessary as part of the maintenance of order against atypical individuals or groups who would disrupt it. When the ability of the political system to provide security for the population breaks down, popular support disappears. Conditions of widespread threat to security are particularly present when food is inaccessible to the population at large. In this case, the underlying reason for support of the system is eliminated, and at the same time there is ``nothing to lose," i.e. even the threat of death does not deter actions that are taken in opposition to the political order. Any incident  then triggers death-defying protests and other actions that disrupt the existing order. Widespread and extreme actions that jeopardize the leadership of the political system, or the political system itself, take place.  All support for the system and allowance for its failings are lost. The loss of support occurs even if the political system is not directly responsible for the food security failure, as is the case if the primary responsibility lies in the global food supply system. 

The role of global food prices in social unrest can be identified from news reports of food riots. Figure \ref{fig:foodprotestsdate} shows a measure of global food prices, the UN Food and Agriculture Organization (FAO) Food Price Index \cite{FAOindex} and the timing of reported food riots in recent years. In 2008 more than 60 food riots occurred worldwide \cite{riotsstats2} in 30 different countries \cite{riotsstats1}, 10 of which resulted in multiple deaths \cite{Cameroon1, Mozambique1, Yemen1, Egypt1, Sudan1, Haiti1, CotedIvoire1, Somalia2, Tunisia1, India2, Sudan2}, as shown in the figure.  
After an intermediate drop, even higher prices at the end of 2010 and the beginning of 2011 coincided with additional food riots (in Mauritania and Uganda \cite{Mauritania2, Uganda1}), as well as the larger protests and government changes in North Africa and the Middle East known as the Arab Spring \cite{Tunisia2, Algeria1, Egypt2, Sudan3, SaudiArabia1, Morocco1, Oman1, Libya1, Yemen2, Iraq1, Bahrain1, Syria1}. There are comparatively fewer food riots when the global food prices are lower. Three of these, at the lowest global food prices, are associated with specific local factors affecting the availability of food: refugee conditions in Burundi in 2005 \cite{Burundi1}, social and agricultural disruption in Somalia \cite{Somalia1} and supply disruptions due to floods in India \cite{India1, India2}. The latter two occurred in 2007 as global food prices began to increase but were not directly associated with the global food prices according to news reports. Two additional food riots in 2007 and 2010, in Mauritania \cite{Mauritania1} and Mozambique \cite{Mozambique2}, occurred when global food prices were high, but not at the level of most riots, and thus appear to be early events associated with increasing global food prices.

These observations are consistent with a hypothesis that high global food prices are a precipitating condition for social unrest. More specifically, food riots occur above a threshold of the FAO price index of $210$ ($p<10^{-7}$, binomial test).
The observations also suggest that the events in North Africa and the Middle East were triggered by food prices. Considering the period of time from January 1990 to May 2011 (Fig. \ref{fig:foodprotestsdate} inset), the probability that the unrest in North Africa and the Middle East occurred by chance at a period of high food prices is 
$p<0.06$ (one sample binomial test). This conservative estimate considers unrest across all countries to be a single unique event over this period of just over twenty years. If individual country events are considered to be independent, because the precipitating conditions must be sufficient for mass violence in each,  
the probability of coincidence is much lower.

\begin{figure}[t]
\centerline{
	\includegraphics[width=150mm]{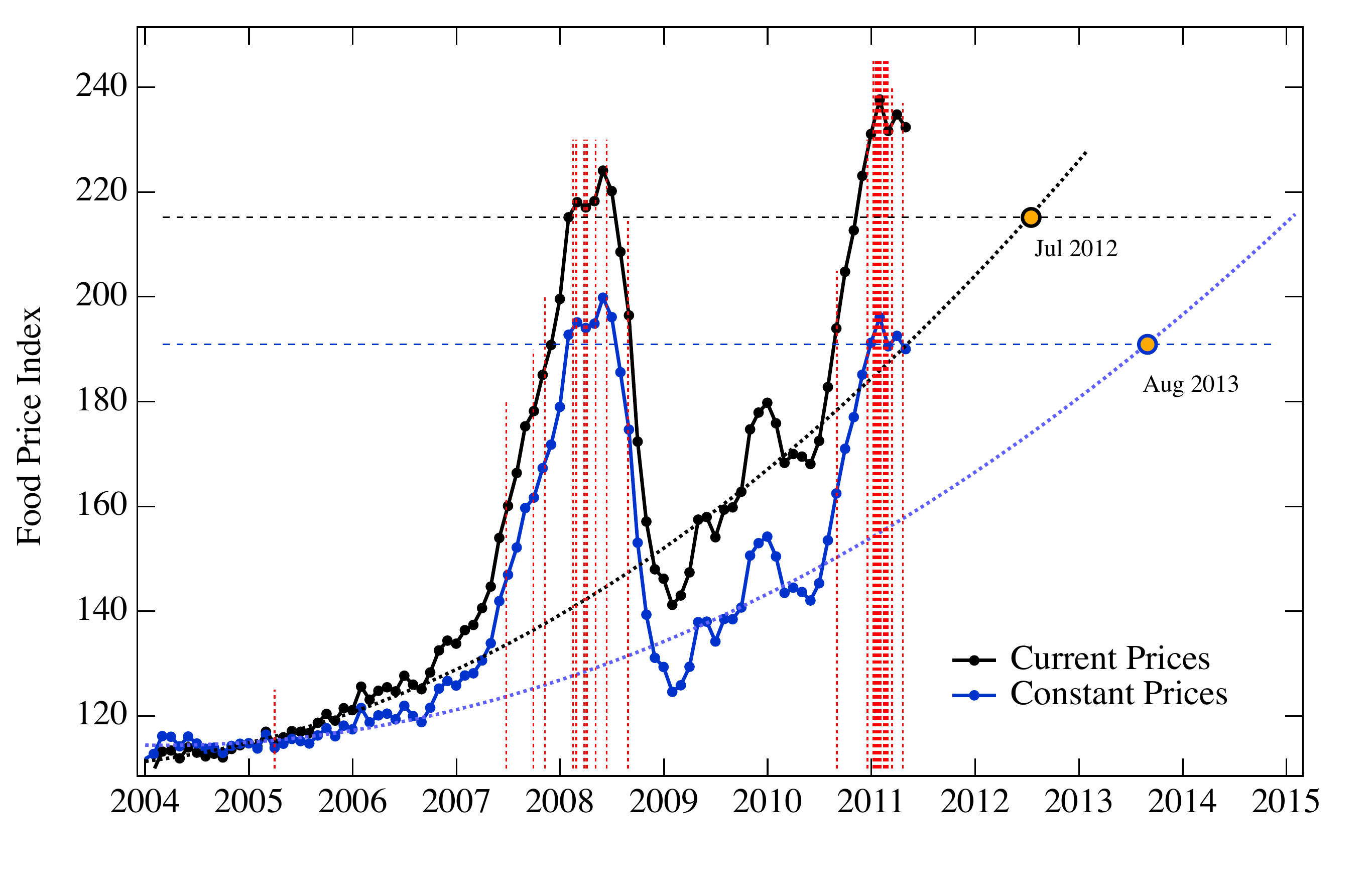}
}
\caption{Time dependence of FAO Price Index at current prices (upper black curve) and constant prices (corrected for inflation, lower blue curve) from January 2004 to May 2011. Red dashed vertical lines correspond to beginning dates of food riots and events associated with the major recent unrest in North Africa and the Middle East. Black and blue horizontal lines represent the price threshold above which riots are ignited in current and constant prices respectively. Index backgrounds are fitted with a third-order polynomial; intersection with the threshold (July 2012 at current prices, August 2013 at prices corrected for world inflation, \cite{World_inflation}) represents the point of instability.\label{fig:extrapolation}}
\end{figure}

A persistence of global food prices above this food price threshold should lead to persistent and increasing global unrest. Given the sharp peaks of food prices we might expect the prices of food to decline shortly. However, underlying the peaks in Fig. \ref{fig:foodprotestsdate}, we see a more gradual, but still rapid, increase of the food prices during the period starting in 2004. It is reasonable to hypothesize that when this underlying trend exceeds the threshold, the security of vulnerable populations will be broadly and persistently compromised. Such a threat to security should be a key concern to policymakers worldwide. Social unrest and political instability of countries can be expected to spread as the impact of loss of security persists and becomes pervasive, even though the underlying causes are global food prices and are not necessarily due to specific governmental policies. While some variation in the form of unrest may occur due to local differences in government, desperate populations are likely to resort to violence even in democratic regimes. A breakdown of social order as a result of loss of food security has been predicted based upon historical events and the expectation that global population increases and resource constraints will lead to catastrophe \cite{Malthus1798, Ehrlich1968, Diamond2005, Tainter2005}. As shown in Fig. \ref{fig:extrapolation}, the underlying trend of increasing prices will reach the threshold of instability in July 2012, if we consider current prices, and April 2013 if we correct prices for reported inflation. Either way, the amount of time until the often warned global food crises appears to be very short. Indeed, consistent with our analysis, the current food price bubble is already subjecting large populations to reported distress, as described in a recent UN report warning of the growing crisis \cite{FAOreport}. 

In a separate paper we consider the causes of the increases in food prices \cite{Lagi_unpublished}. While there have been several suggested origins of the food price increases, we find the dominant ones to be investor speculation and ethanol production. Our analysis shows that the two parts of the dynamics of prices can be directly attributed to the two different causes: the price peaks are due to speculators causing price bubbles, and the background increase shown in Fig. \ref{fig:extrapolation} is due to corn to ethanol conversion. This intuitive result is made quantitative by the analysis in that paper. 

Both factors in food prices can be linked directly to recent US governmental actions. Speculator activity has been enhanced by deregulation of the commodities markets that exempted dealers from trading limits \cite{Marco-article,FedReg,Stewart2008}, and subsidies and other policies have been central to the growth of ethanol conversion \cite{Hahn2009,Rismiller2009}.

The importance of food prices for social stability points to the level of human suffering that may be caused by increased food prices. The analysis we presented of the timing of peaks in global food prices and social unrest implies that the 2011 unrest was precipitated by a food crisis that is threatening the security of vulnerable populations. Deterioration in food security led to conditions in which random events trigger widespread violence. The condition of these vulnerable populations could have been much worse except that some countries controlled food prices in 2011 due to the unrest in 2008 \cite{Egypt_control1,Egypt_control2,Cameroon_control,China_control,Ukraine_control,Bangladesh_control,Vietnam_control,Tajikistan_control,Pakistan_control,TandT_control,Rwanda_control,Indonesia_control}. Food price controls in the face of high global food prices carry associated costs. Because of the strong cascade of events in the Middle East and North Africa only some countries had to fail to adequately control food prices for events to unfold \cite{Algeria_no_control,Tunisia_no_control,Tunisia_Algeria_no_control,Egypt_no_control,All_no_control}. This understanding suggests that reconsidering biofuel policy as well as commodity market regulations should be an urgent priority for policymakers. Reducing the amount of corn converted to ethanol, and restricting commodity future markets to bona fide risk hedging would reduce global food prices \cite{Lagi_unpublished}. The current problem transcends the specific national political crises to represent a global concern about vulnerable populations and social order. 

Our analysis of the link between  global food prices and social unrest supports a growing conclusion that it is possible to build mathematical models of global economic and social crises \cite{Lim2007,MacKenzie2011Mar,Pozen2008,Harmon2008,Vedantam2009,Harmon2010Nov,Keim2010,Harmon2010Feb,Keim2011,Kelland2011}. Identifying a signature of unrest for future events is surely useful. Significantly, prior to the unrest, on December 13, 2010, we submitted a government report \cite{NECSIreport} analyzing the repercussions of the global financial crises, and directly identifying the risk of social unrest and political instability due to food prices (see Fig. \ref{fig:foodprotestsdate}). This report, submitted four days before the initial human trigger event, the action of Mohamed Bouazizi in Tunisia \cite{Abouzeid2011,Watson2011}, demonstrates that it is possible to identify early warning signs before events occur. Prediction is a major challenge for socio-economic analysis. Understanding  when and whether prediction is possible is important for science and policy decisions. Our predictions are conditional on the circumstances, and thus allow for policy interventions to change them. Whether policy makers will act depends on the various pressures that are applied to them, including both the public and special interests. 

We thank Blake Stacey, Ama\c{c} Herda\u{g}delen, Alexander Gard-Murray, Andreas Gros, and Shlomiya Bar-Yam for helpful comments on the manuscript. This work was supported in part by AFOSR under grant FA9550-09-1-0324 and ONR under grant N000140910516.

\bibliographystyle{Science}

\end{document}